\documentclass{pasj00}

\begin{document}
\SetRunningHead{T. Kato, R. Ishioka, and M. Uemura}{Photometric Study of KR Aur}

\Received{}
\Accepted{}

\title{Photometric Study of KR Aurigae during the High State in 2001}

\author{Taichi \textsc{Kato}, Ryoko \textsc{Ishioka}, Makoto \textsc{Uemura}}
\affil{Department of Astronomy, Kyoto University,
       Sakyou-ku, Kyoto 606-8502}
\email{(tkato,ishioka,uemura)@kusastro.kyoto-u.ac.jp}


\maketitle

\begin{abstract}
  We photometrically observed the VY Scl-type cataclysmic variable
KR Aurigae after its final rise from the fading episode in 2000--2001.
Time-resolved observation revealed that the light curve is dominated by
persistent short-term variation with time-scales of minutes to tens of
minutes.  On some nights, quasi-periodic variations with periods of 10--15 min
were observed.  No coherent variation was detected.  The power spectral
density of the variation has a power law component ($f^{-1.63}$).
The temporal properties of the short-term variations in KR Aur present
additional support for the possibility that flickering in CVs may be
better understood as a result of self-organized critical state as in
black-hole candidates.  The light curve lacks ``superhump"-type signals,
which are relatively frequently seen in VY Scl-type systems and which are
suggested to arise from tidal instability of the accretion disk induced
by changing mass-transfer rates.  The present observation suggests
a borderline of superhump excitation in VY Scl-type stars between mass
ratios $q$=0.43 (MV Lyr) and $q$=0.60 (KR Aur).
\end{abstract}

\section{Introduction}

   Cataclysmic variables (CVs) are close binary systems consisting of
a white dwarf and a red dwarf secondary transferring matter via the Roche
lobe overflow.  The resultant accretion disk becomes thermally stable
in systems with high mass-transfer rates ($\dot{M}$).  Such systems are
called novalike (NL) stars, because they lack outbursts characteristic
to dwarf novae [see \citet{osa96review} for a review].  Among NL stars,
there exists a small group which shows a temporary reduction or
cessation of $\dot{M}$ from the secondary.  These systems are called
VY Scl-type stars or anti-dwarf novae (\cite{war95book}).

   Although accretion disks in NL stars are thermally stable, the
disk can be tidally unstable (\cite{osa95eruma}; \cite{osa96review}).
Numerical simulations have shown that this instability (tidal instability:
\cite{whi88tidal}) only appears below a certain mass ratio ($q=M_2/M_1$):
$q<0.25-0.33$ depending on calculations (\cite{hir90SHexcess};
\cite{woo00SH}; \cite{whi91SH}; \cite{mol92SHexcess}; \cite{mur98SH}).
In recent years, several systems above this stability limit are known
to show superhumps \citep{pat99SH}.  Since many of them are VY Scl-type
stars, there has been a theoretical interpretation that accretion disks
can be tidally unstable upon variation of $\dot{M}$ \citep{mur00SHvyscl}
even in intermediate $q$ systems.  Temporary appearance of superhump
signals in a recent low state of a VY Scl-type star, BZ Cam
\citep{kat01bzcam} may support this interpretation.
However, observations have not yet fully illustrated
the upper $q$ limit for superhumpers in VY Scl-type stars.
KR Aur ($q$=0.60) is an ideal system to examine such a condition,
since this system has the smallest $q$ among VY Scl-type stars
which have not been reported to show (or studied for) superhumps.

\begin{figure*}
  \begin{center}
    \FigureFile(160mm,80mm){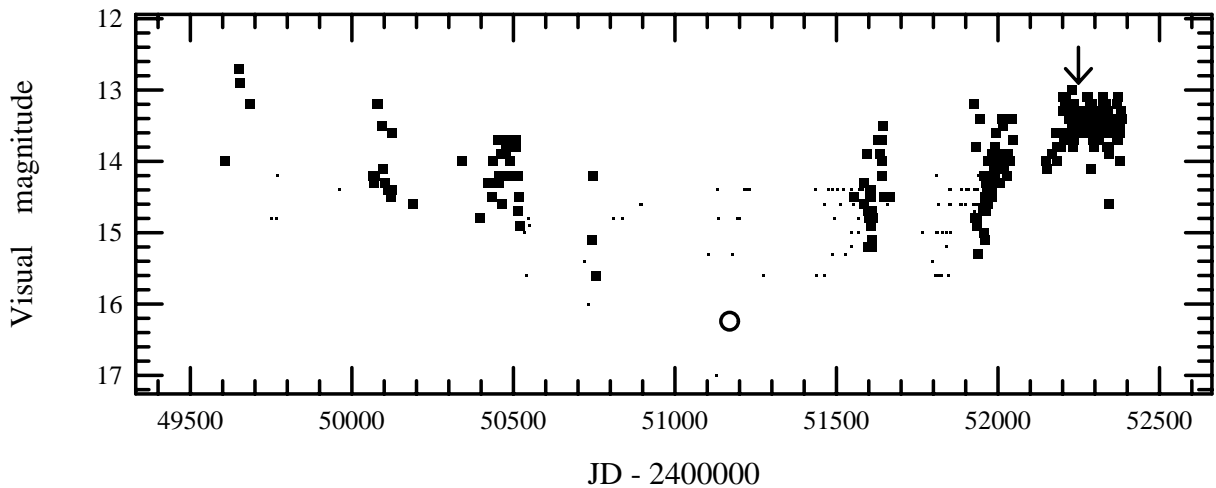}
  \end{center}
  \caption{Long-term light variation of KR Aur between 1994 and 2002
  drawn from visual observations reported to VSNET.
  Large dots and small dots represent positive and negative (upper limit:
  object undetected) observations, respectively.  Typical errors of
  visual observations are 0.2--0.3 mag.
  The open circle represents our snapshot CCD observations.
  The mean epoch of our time-series CCD photometry is shown by an arrow.
  }
  \label{fig:vis}
\end{figure*}

\section{Past Studies of KR Aurigae}

   KR Aur is a variable star originally discovered by M. Popova in 1960.
Popova originally classified the object as a `unusual nova' (cf.
\cite{due87novaatlas}).  The object has received much attention since
\citet{pop77kraur} claimed, from the detection of a large infall
velocity of 3200 km s$^{-1}$ in its H$\beta$ line, that the object may
be a solitary black hole accreting interstellar matter.
\citet{pop78kraur} further reported a long-term light curve, which
could not fit any existing classifications of variable stars, a large
radial velocity (+130 km s$^{-1}$), a steep Balmer decrement, and
a spectral energy distribution approximated by $F_{\nu}$=const.
\citet{pop78kraur} considered that these features are also consistent
with an accreting black hole.  However, \citet{dor77kraurspec} reported
that time-scales of continuum variation (a few hours to a few tens of
seconds) and that $UBV$ colors resemble those of CVs, known as ``novalike"
stars at the time of the reporting.

   \citet{dor78kraur} studied short-term light variation, and detected
$\sim$0.4 mag variations with time-scales of several minutes.
\citet{lil80kraur} studied archival plates, and obtained the first
long-term light curve of this object.  \citet{lil80kraur} showed that
KR Aur sometimes show excursions to faint states (which was also noted
by \citet{pop78kraur}), but no clear periodicity was found.
\citet{pop82kraur} reported an ongoing new fading episode.

   The nature of the object was revealed with spectroscopic observations
by \citet{sha83kraur} and \citet{hut83kraurorbit}, who detected a binary
motion from radial velocity study.  The object was thus confirmed to be
a CV.  \citet{wil83CVspec1} showed that the object shows a spectrum
typical of a CV with relatively weak emission lines.
However, since the characteristics of light variations did not fit any
known subclass of CVs, long-term light variations were
subsequently studied in detail mainly from photographic survey materials
(\cite{got83kraur}; \cite{got84kraur}; \cite{got85kraur}; \cite{got86kraur};
\cite{got87kraur}; \cite{got88kraur}; \cite{got89kraur}; \cite{got90kraur};
\cite{fuj87kraur}; \cite{lil84kraur}; \cite{pop94kraur}; \cite{ant96kraur}).
All of these observations showed rather erratic light curves, and occasional
fadings with time-scales of months to a few years.  From these
characteristics, KR Aur is now considered to belong to VY Scl-type novalike
CVs, which occasionally show fading episodes (cf. \cite{war74vysclzcam};
\cite{gar88vyscldqher}; \cite{gre98vyscl}; \cite{lea99vyscl}).
KR Aur was also independently selected as an ultraviolet-excess object
(KUV 06126+2836: \cite{kon84KUV2}), for which \citet{weg90KUVspec}
gave a spectroscopic classification as a CV.

   \citet{pop89kraur} reported rapid variations in KR Aur.
\citet{bir90kraur} further studied this star and reported the presence
of 25-min periodicity.  The light curve presented by \citet{bir90kraur}
suggests a moderate degree of coherence.

\begin{figure}
  \begin{center}
    \FigureFile(88mm,60mm){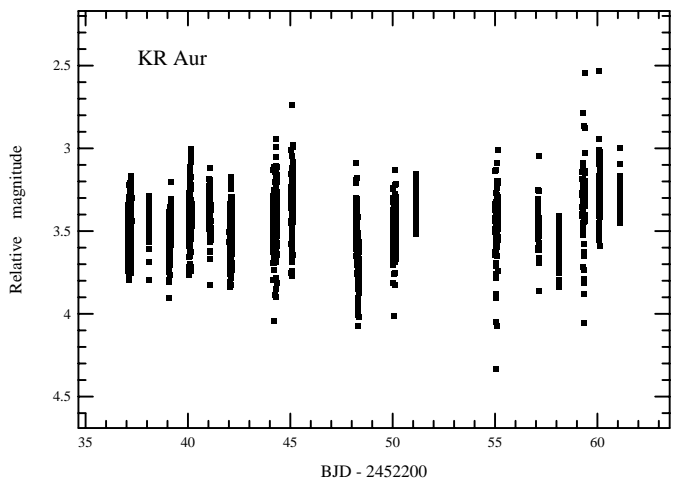}
  \end{center}
  \caption{Overall light variation of KR Aur in 2001 November--December.
  }
  \label{fig:overall}
\end{figure}

\begin{table*}
\caption{Log of observations.}\label{tab:log}
\begin{center}
\begin{tabular}{lccccc}
\hline\hline
Date      & BJD$^*$ (start--end) & N$^\dagger$ & Mag$^\ddagger$ &
            Error$^\S$ & Inst$^\|$ \\
\hline
1998 December 21 & 51169.248--51169.248 &   3 & 6.408 & 0.036 & 1 \\
2001 November 23 & 52237.059--52237.244 & 428 & 3.466 & 0.006 & 2 \\
2001 November 24 & 52238.082--52238.111 &  69 & 3.464 & 0.010 & 2 \\
2001 November 25 & 52239.055--52239.172 & 270 & 3.526 & 0.006 & 2 \\
2001 November 26 & 52240.042--52240.177 & 252 & 3.376 & 0.008 & 2 \\
2001 November 27 & 52241.041--52241.105 & 142 & 3.369 & 0.008 & 2 \\
2001 November 28 & 52242.029--52242.152 & 259 & 3.508 & 0.008 & 2 \\
2001 November 30 & 52244.149--52244.352 & 355 & 3.427 & 0.007 & 2 \\
2001 December  1 & 52245.050--52245.126 & 102 & 3.320 & 0.018 & 2 \\
2001 December  4 & 52248.211--52248.346 & 281 & 3.595 & 0.010 & 3 \\
2001 December  6 & 52250.007--52250.168 & 359 & 3.465 & 0.006 & 2 \\
2001 December  7 & 52251.116--52251.152 &  61 & 3.285 & 0.011 & 3 \\
2001 December 11 & 52255.003--52255.166 & 200 & 3.455 & 0.010 & 2 \\
2001 December 13 & 52257.104--52257.137 &  51 & 3.460 & 0.017 & 3 \\
2001 December 14 & 52258.109--52258.134 &  53 & 3.614 & 0.015 & 3 \\
2001 December 15 & 52259.261--52259.073 &  94 & 3.308 & 0.022 & 3 \\
2001 December 16 & 52260.073--52260.128 &  91 & 3.229 & 0.018 & 3 \\
2001 December 17 & 52261.092--52261.126 &  69 & 3.272 & 0.009 & 3 \\
\hline
 \multicolumn{6}{l}{$^*$ BJD$-$2400000.} \\
 \multicolumn{6}{l}{$^\dagger$ Number of frames.} \\
 \multicolumn{6}{l}{$^\ddagger$ Averaged magnitude relative to GSC 1889.700.} \\
 \multicolumn{6}{l}{$^\S$ Standard error of the averaged magnitude.} \\
 \multicolumn{6}{l}{$^\|$ 1: Kyoto (25-cm + ST-7),
                              2: Kyoto (25-cm + ST-7E).} \\
 \multicolumn{6}{l}{\phantom{$^\|$} 3: Kyoto (30-cm + ST-7E).} \\
\end{tabular}
\end{center}
\end{table*}

\section{Observations}

    The observations were mainly acquired using an unfiltered ST-7E camera
attached to 25-cm/30-cm Schmidt-Cassegrain telescopes at Kyoto University.
A single snapshot observation during a low state was taken on 1998
December 21, using an unfiltered ST-7 camera attached to a 25-cm
Schmidt-Cassegrain telescope.  All systems give magnitudes close to
$R_{\rm c}$.  The images were dark-subtracted, flat-fielded, and analyzed
using the Java$^{\rm TM}$-based aperture photometry package developed by
one of the authors (TK).  The differential magnitudes of the
variable were measured against GSC 1889.700 (Tycho-2 magnitude
$V$=10.27, $B-V$=+0.88), whose constancy during the run was confirmed by
comparison with fainter check stars in the same field.
The log of observations is summarized in table \ref{tab:log}.
The total number of useful frames was 3139.
Barycentric corrections were applied before the following analysis.

\begin{figure}
  \begin{center}
    \FigureFile(88mm,60mm){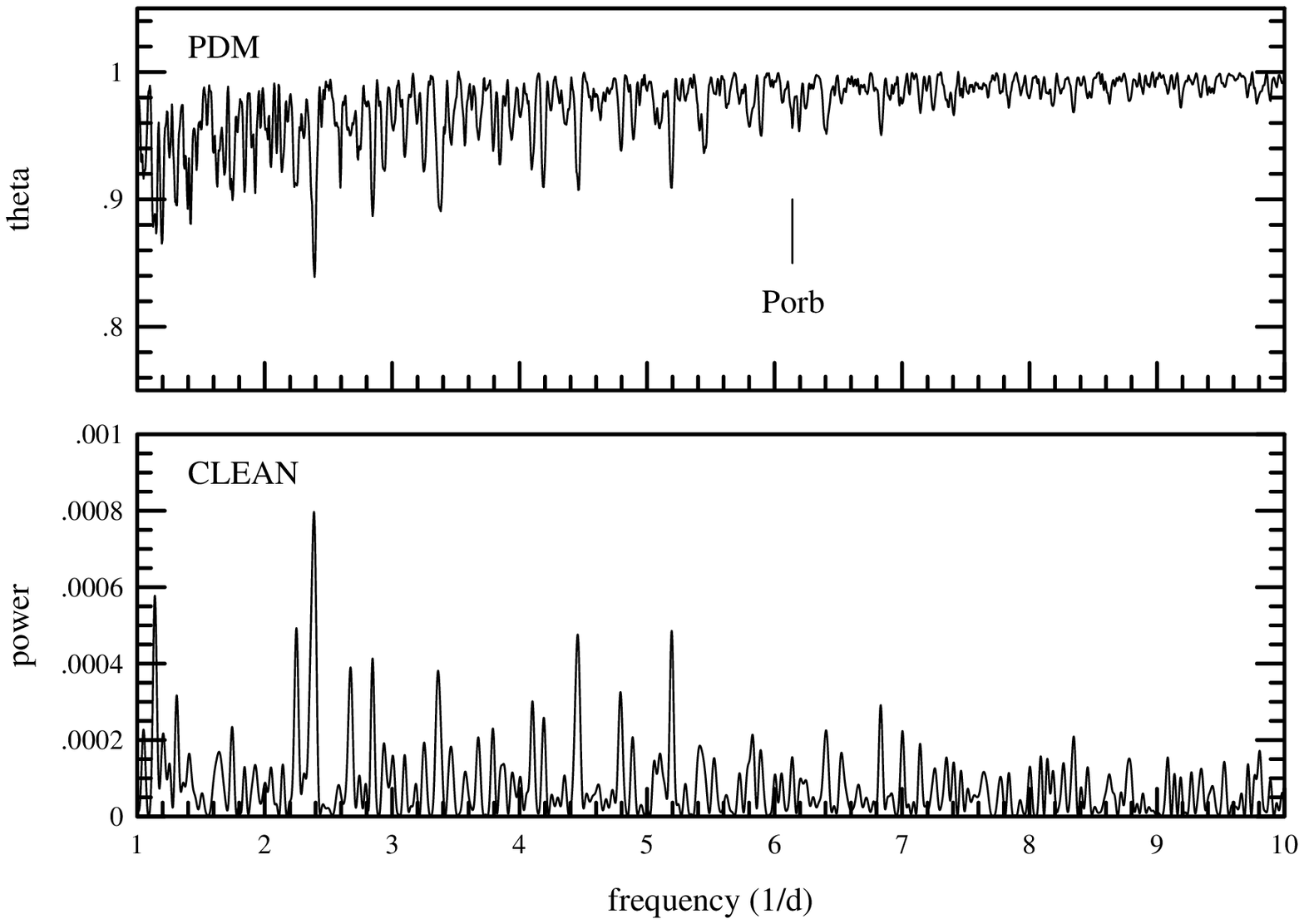}
  \end{center}
  \caption{Period analysis of the November 23--December 17 data.  The upper
  and lower panels represent the results of period analysis with Phase
  Dispersion Minimization \citep{PDM} and with the CLEAN algorithm
  \citep{CLEAN}, respectively.
  The orbital period is marked with a tick on the upper panel.
  No significant coherent periodicity was detected at or near the
  orbital period.
  }
  \label{fig:perana}
\end{figure}

\begin{figure*}
  \begin{center}
    \FigureFile(130mm,90mm){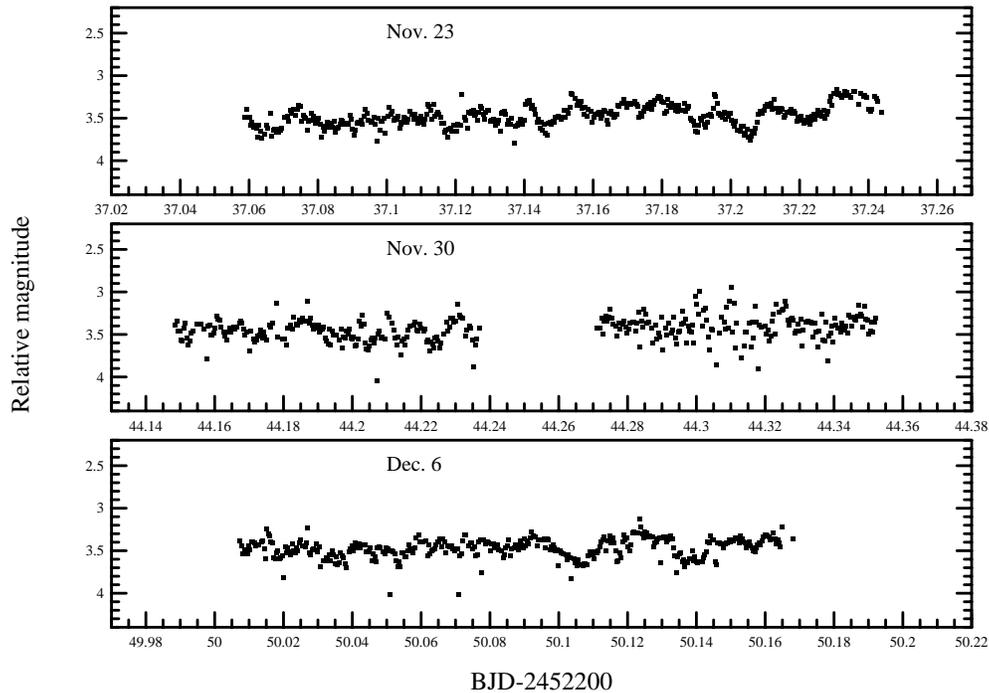}
  \end{center}
  \caption{Representative nightly light curves at different epochs.
  All light curves show distinct short-term (minutes to tens of minutes)
  quasi-periodic variations.  No systematic variation close to the orbital
  period (0.16280 d) was observed (cf. subsection \ref{sec:sh}).
  }
  \label{fig:nightly}
\end{figure*}

\begin{figure}
  \begin{center}
    \FigureFile(88mm,120mm){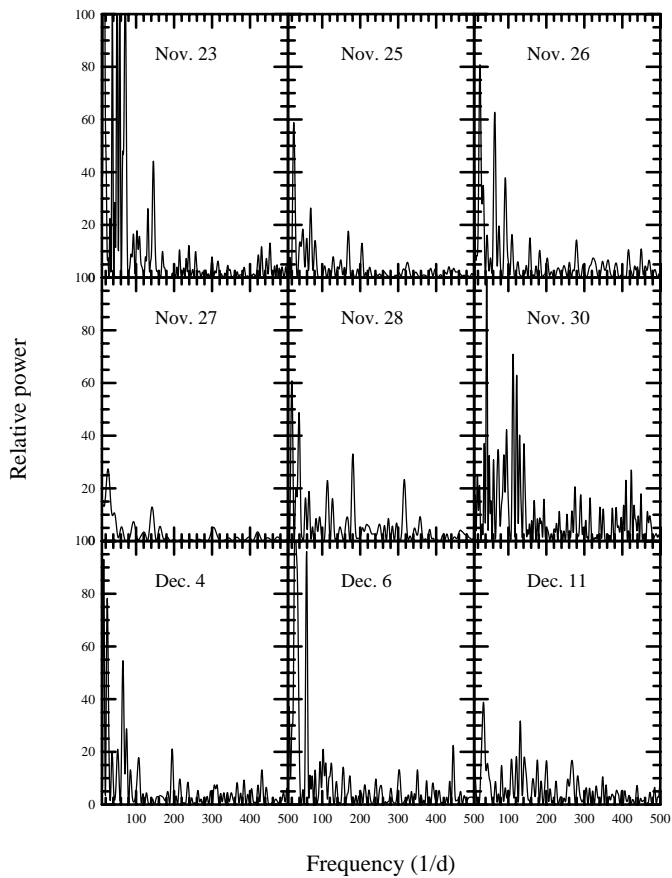}
  \end{center}
  \caption{Power spectra of nightly observations.  On some nights
  (November 23, 30, December 11), increased power around the frequencies
  100--150 d$^{-1}$ (corresponding to the periods of 10--15 min) was
  observed.  There was no common period between the nightly observations.
  }
  \label{fig:power}
\end{figure}

\section{Results and Discussions}

\subsection{Long-Term Variation}

  Figure \ref{fig:vis} long-term shows light variation of KR Aur between
1994 and 2002 drawn from visual observations reported to
VSNET Collaboration.\footnote{
$\langle$ http://www.kusastro.kyoto-u.ac.jp/vsnet/ $\rangle$.
}  Typical errors of visual observations are 0.2--0.3 mag, which will not
affect the following analysis.
The light curve shows predominant low states (durations: several
months to two years) when the object is fainter than mag 14, and occasional
high states when the object is typically between mag 13 and 14.
Such a high occurrence of low states is relatively rare among VY Scl-type
stars \citep{gre98vyscl}.  The long-term behavior of the system most resembles
``superminimum" of MV Lyr (\cite{rob81mvlyr}; \cite{wen83mvlyr};
\cite{fuh85mvlyr}).  At a closer look, the system underwent a small
dwarf nova-like outburst around JD 2450745--2450755.  The object then
became inactive.  Before the fully recovery, the object underwent a
rather complex brightening around JD 2451580--2451630.  This behavior is
quite reminiscent of the fading and recovery processes observed in
MV Lyr (\cite{pav98mvlyr}; \cite{shu98mvlyr}; \cite{pav99mvlyr}).
Although such behavior is unexpected for variable mass-transfer rates
on a usual CV \citep{hon94v794aql}, \citet{lea99vyscl} showed that,
in the presence of heating from a hot white dwarf, the irradiation
on the accretion disk suppresses the thermal instability, which can
reproduce the observed light curve of VY Scl-type systems.  From these
observations, we propose that MV Lyr and KR Aur comprise the most
``active" subgroup of VY Scl-type stars.

  Figure \ref{fig:overall} presents overall light variation of KR Aur in
2001 November--December drawn from observations in table \ref{tab:log}.
Although some irregular variation is superimposed,
no major ``flare"-like brightening (cf. \cite{pop78kraur}) was observed.

\subsection{Superhumps}\label{sec:sh}

   Figure \ref{fig:perana} presents a period analysis of the
November 23--December 17 data, after subtracting a linear fit to the overall
light curve.  The upper and lower panels represent
the results of period analysis with Phase Dispersion Minimization \citep{PDM}
and with the CLEAN algorithm \citep{CLEAN}, respectively.  The orbital period
is marked with a tick on the upper panel.  No significant coherent
periodicity was detected at or near the orbital period, indicating that
no detectable periodic or quasi-periodic variations related to
orbital modulations or superhumps were present (see also subsection
\ref{sec:short}).

   The absence of superhumps makes a clear contrast to other VY Scl-type
stars with superhumps (TT Ari: \citet{uda88ttari}; \citet{ski98ttari};
\citet{and99ttari}; \citet{kra99ttari}; \citet{sta01ttariSHspec},
MV Lyr: \citet{ski95mvlyr}, V751 Cyg: \citet{pat01v751cyg}).
This finding suggests that the mechanism proposed by \citet{mur00SHvyscl}
is ineffective in the mass ratio of KR Aur.
We propose that there is a borderline of
superhump excitation between mass ratios $q$=0.43 (MV Lyr) and $q$=0.60
(KR Aur) in VY Scl-type stars.  Future determination of mass ratios
in longer $P_{\rm orb}$ systems (i.e. candidates for systems with higher
$q$ than MV Lyr) BZ Cam, V751 Cyg and TT Ari is expected to provide
a more stringent constraint to this limit (see table \ref{tab:vyscl}
for a summary of superhumps and binary parameters of VY Scl-type stars).

\begin{table}
\caption{Parameters of VY Scl-Type Stars$^*$.}\label{tab:vyscl}
\begin{center}
\begin{tabular}{llccc}
\hline\hline
Object & $P_{\rm orb}$ (d) & $P_{\rm SH}$ (d) & $M_1$ & $q$ \\
\hline
VY Scl      & 0.232    & $\cdots$ & 1.22 & 0.32 \\
KR Aur      & 0.16280  & $\cdots$ & 0.59 & 0.60 \\
LX Ser$^\dagger$ & 0.158432 & $\cdots$ & 0.41 & 0.88 \\
BZ Cam      & 0.153693 & 0.15634  & $\cdots$ & $\cdots$ \\
V794 Aql    & 0.1533   & $\cdots$ & $\cdots$ & $\cdots$ \\
V425 Cas    & 0.1496   & $\cdots$ & $\cdots$ & $\cdots$ \\
VZ Scl      & 0.144622 & $\cdots$ & 1:   & 0.7: \\
V751 Cyg    & 0.144464 & 0.1394   & $\cdots$ & $\cdots$ \\
PG 1000+667 & 0.144384 & $\cdots$ & $\cdots$ & $\cdots$ \\
TT Ari      & 0.137551 & 0.133160 & $\cdots$ & $\cdots$ \\
            &          & 0.148815 &      &      \\
DW UMa$^\dagger$ & 0.136607 & 0.1330   & 0.9  & 0.32 \\
MV Lyr      & 0.1329   & 0.138    & $\cdots$ & 0.43 \\
V442 Oph$^\dagger$ & 0.124330 & 0.12090  & $\cdots$ & $\cdots$ \\
\hline
\end{tabular}
\end{center}
{\footnotesize
  $^*$ LQ Peg \citep{kat99lqpeg} is also known as a VY Scl-type star.
The orbital period has not been reported.

  $^\dagger$ SW Sex star (see \cite{hel00swsexreview} for a recent
review; see also \cite{tho91pxand}).  A few other SW Sex stars
[PX And (\cite{sti95pxand}); BH Lyn (\cite{hoa97bhlyn})]
are claimed to show some degree of low/high state transitions
(mainly from spectroscopic observations).
Such a variation may more represent a change in
the disk state (cf. \cite{gro01swsex}) rather than VY Scl-type
temporary reduction or cessation of $\dot{M}$ from the secondary.
Some DQ Her stars (intermediate polars) also show temporary fadings
(\cite{gar88vyscldqher}; \cite{hes00CVlowstate}), but they are not
usually considered as VY Scl-type stars.

{\bf References:}
  VY Scl: \citet{mar00vyscl};
  KR Aur: \citet{sha83kraur};
  LX Ser: \citet{you81lxser};
  BZ Cam: \citet{pat96bzcam}, \citet{kat01bzcam};
  V794 Aql: \citet{hon98v794aql};
  V425 Cas: see \citet{kat01v425cas} and the references therein;
  VZ Scl: \citet{war75vzscl}, \citet{rob76CVmass}, \citet{odo87vzscl},
          \citet{she84vzscl};
  V751 Cyg: \citet{pat01v751cyg};
  PG 1000+667: \citet{hil98pg1000};
  TT Ari: \citet{cow75ttari}, \citet{tho85ttari}, \citet{uda88ttari},
          \citet{ski98ttari}, \citet{and99ttari}, \citet{kra99ttari},
          \citet{sta01ttariSHspec};
  DW UMa: \citet{sha88dwuma}, \citet{pat99SH}, \citet{bir00dwuma};
  MV Lyr: \citet{ski95mvlyr};
  V442 Oph: \citet{pat99SH}, \citet{hoa00v442oph}, \citet{dia01v442oph}.
}
\end{table}

\subsection{Short-Term Variations}\label{sec:short}

   Figure \ref{fig:nightly} shows typical examples of nightly light curves.
Three longest runs (November 23 and 30, December 6) were selected as
representatives of different epochs of the present observation.
All light curves show distinct short-term (minutes to tens of minutes)
quasi-periodic variations.  No systematic variation close to the orbital
period (0.16280 d) was observed (cf. subsection \ref{sec:sh}).
Aside from the lack of superhumps, these variations look remarkably
similar to those of quasi-periodic oscillations (QPOs) observed in another
VY Scl-type star, TT Ari (e.g. \cite{mar80ttari}; \cite{jen83ttari};
\cite{hol92ttariQPO}; \cite{tre96ttari}; \cite{and99ttari}).

   Figure \ref{fig:power} shows power spectra of nightly observations.
On some nights (November 23, 30, December 11), an increased power around the
frequencies 100--150 d$^{-1}$ (corresponding to the periods of 10--15 min) was
observed.  There was no common period between the nightly observations.
This finding confirms the quasi-periodic nature of the short-term variations.
We have not been able to confirm the presence of 25-min periodicity claimed
by \citet{bir90kraur}.  The periods of the presently observed QPOs are
close to the periodicities (480--780 s) recorded by \citet{sin93kraurQPO}.
We have also confirmed night-to-night variation of dominant periods
as was originally claimed by \citet{sin93kraurQPO}.

   We also note that the overall profiles of short-term variations
(figure \ref{fig:nightly}) and the nightly variation of the power spectra
are also similar to those observed in the peculiar symbiotic binary
V694 Mon (\cite{mic93v694mon}; \cite{dob96v694mon}; \cite{ish01v694mon}).

   There are two major types of ``quasi-periodic" oscillations in
CVs: dwarf nova oscillations (DNOs), which are oscillations observed only
in dwarf nova outbursts, having periods of 19--29~s, and have long
(several tens to $\sim$100 wave numbers) coherence times (\cite{rob73DNO};
\cite{szk76DNO}; \cite{hil80DNO}), and (in a narrower sense)
QPOs (for a review, see \citet{war95book}).  The present QPOs in KR Aur
correspond to the latter classification.  Several models have been proposed
to account for QPOs, including vertical or radial oscillations of the
accretion disk \citep{kat78QPO}, reprocessing of the light by the orbiting
blobs \citep{pat79aeaqr}, non-radial pulsations of the accretion disk
(\cite{pap78CVQPO}; \cite{vanhor80DNQPO}), radial oscillation of the
accretion disk (\cite{cox81QPO}; \cite{blu84QPO}; \cite{oku91QPO};
\cite{oku92QPO}), excitation of trapped oscillations around the discontinuity
of physical parameters \citep{yam95DNoscillation}.  There has been also
a suggestion that the magnetism of the white dwarf can be responsible
for some kinds of QPOs \citep{mik90chcygflickering}.  Although present
observations are not able to constrain the origin of QPOs, a future search
for coherent X-ray, ultraviolet or optical pulsations, which are a well-known
signature of a magnetic white dwarf, would be helpful in discriminating
the possibilities.

\begin{figure}
  \begin{center}
    \FigureFile(88mm,60mm){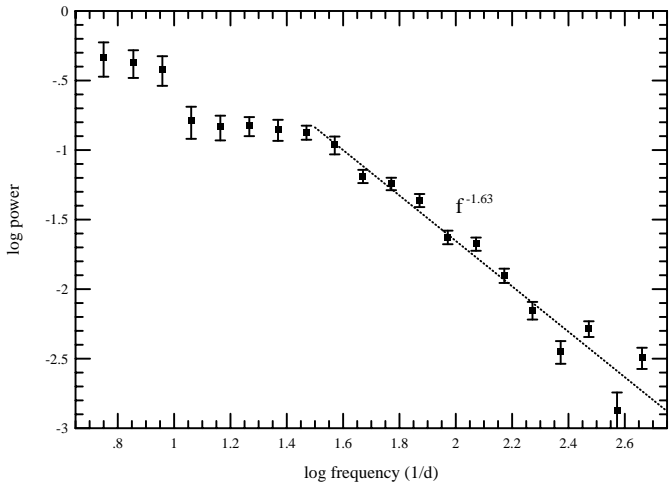}
  \end{center}
  \caption{Power spectral density (PSD) of the entire data set.
  High-frequency white noise has been subtracted from the PSD.
  Above the frequency $\log f$(d$^{-1}$)=1.5, the PSD is proportional
  to $f^{-1.63}$.
  }
  \label{fig:powave}
\end{figure}

\subsection{Power Spectrum of Flickering}

   Figure \ref{fig:powave} shows power spectral density (PSD) of the
entire data set.  High-frequency white noise has been subtracted from the
PSD.  Above the frequency $\log f$=1.5 d$^{-1}$ (corresponding to
time-scales shorter than $\sim$45 min), the PSD is proportional to
$f^{-1.63}$.

   These short-term variations (flickering) are one of the most
characteristic features in CVs.  Although the existence of flickering
in CVs has been well-documented since the 1940's
(see \cite{bru92CVflickering} for an extensive historical review), their
physical origin has not been well understood.  Historically,
\citet{war71ugem} demonstrated that flickering almost disappeared during
eclipses of the eclipsing dwarf nova U Gem.  This finding indicated that
the origin of flickering is strongly associated with the hot spot
(the stream impact point on the accretion disk).  \citet{jam87spotflickering}
noted the presence of power law-type frequency dependence, and proposed
that a multiple scattering from the flickering source (hot spot) is
responsible for this frequency dependence.  More recently,
\citet{bru91NLflickering} more extensively studied the properties of
flickering, and summarized an observational review \citep{bru92CVflickering}.
From the analysis of frequency and color dependencies,
\citet{bru92CVflickering} and \citet{bru93CVboundarylayer} suggested that
the inner part of the accretion disk (rather than the hot spot) is more
responsible for flickering.  More direct observational evidence for
a major contribution from the inner accretion disk to flickering has been
demonstrated through eclipse observations of CVs
(HT Cas: \cite{wel95htcas}; \cite{bru00htcasv2051ophippeguxumaflickering},
Z Cha: \cite{bru96zchaflickering}).
It is now widely believed that the originally proposed idea of stream
impact-type flickering \citep{war71ugem} applies to a only limited
sample of CVs, or contributes to a small extent to overall flickering
activity, and that most of CVs have a strong concentration of flickering
activity toward the inner accretion disk \citep{bru96zchaflickering}.

   In order to reproduce this power-law spectrum, \citet{yon97CVflickering}
proposed a superposition of ``shots" in a self-organized critical state
(SOC), which was originally introduced to explain time variations in
black-hole candidates (BHCs)
(e.g. \cite{min94BHADSOC}; \cite{min94BHfluctuation};
\cite{tak95BHADfluctuation}; \cite{kaw00BHADfluctuation}).
Taking this analogy into account, the power law-type temporal properties
of the short-term variations in KR Aur present additional support for the
possibility that flickering in CVs may be better understood as a result of
SOC as in BHCs.  Although the detailed mechanism of energy release
was not identified at the time of \citet{yon97CVflickering},
\citet{wil02ADflare} recently tried to explain flickering by a
superposition of flares, resulting from an energy release in the photosphere
of the accretion disk of injected high-energy electrons originating from
reconnections of magnetic field lines.  This type of energy release in visual
wavelengths would be a promising candidate for explaining flickering
in CVs.  Further quantitative comparisons with numerical simulations
and observed properties in CVs will be a next step toward understanding
flickering in CVs.

\section{Summary}

  We photometrically observed the VY Scl-type cataclysmic variable
KR Aurigae after its final rise from the fading episode in 2000--2001.
We show that the long-term light curve of KR Aur is densely populated
with low states, making KR Aur an exceptionally active VY Scl-type star.
The object showed a complex recovery process from a faint state,
which may be understood as the result from heating on the accretion disk
from a hot white dwarf.  Time-resolved observation during the high state
revealed that the light curve is dominated by persistent short-term
variation with time-scales of minutes to tens of minutes.  On some nights,
quasi-periodic variations with periods of 10--15 min
were observed.  No coherent variation was detected.  The power spectral
density of the variation has a power law component ($f^{-1.63}$)
at high frequencies.
The temporal properties of the short-term variations in KR Aur present
additional support for the possibility that flickering in CVs may be
better understood as a result of self-organized critical state as in
black-hole candidates.  Contrary to the large-amplitude short-term variations,
the light curve lacks ``superhump"-type signals, which are relatively
frequently seen in VY Scl-type systems.  The present observation suggests
a borderline of superhump excitation in VY Scl-type stars between
mass ratios $q$=0.43 (MV Lyr) and $q$=0.60 (KR Aur).

\vskip 3mm

The authors are grateful to observers who reported vital observations
to VSNET.
This work is partly supported by a grant-in aid (13640239) from the
Japanese Ministry of Education, Culture, Sports, Science and Technology.
Part of this work is supported by a Research Fellowship of the
Japan Society for the Promotion of Science for Young Scientists (MU).

\end{document}